\documentclass[runningheads]{llncs}
\usepackage[T1]{fontenc}
\usepackage{graphicx}
\usepackage{xcolor}
\usepackage{tabularx}
\usepackage{adjustbox}
\usepackage[colorlinks]{hyperref}
\usepackage[colorinlistoftodos]{todonotes}
\usepackage{hyperref}

\begin{document}
%

\title{A Reference Architecture for Gamified Cultural Heritage Applications Leveraging Generative AI and Augmented Reality}

\titlerunning{A Reference Architecture for GAM CH Applications leveraging GAI and AR}

\author{Federico Martusciello\inst{1} \and
Henry Muccini \inst{1}\and 
Antonio Bucchiarone \inst{1}}

\authorrunning{F. Author et al.}

\institute{DISIM, University of L'Aquila
\\
\url{https://www.univaq.it} 
\\
\email{federico.martusciello@graduate.univaq.it, henry.muccini@univaq.it, antonio.bucchiarone@univaq.it}
}
\maketitle              

\begin{abstract}

The rapid advancement of Information and Communication Technologies is transforming Cultural Heritage access, experience, and preservation. However, many digital heritage applications lack interactivity, personalization, and adaptability, limiting user engagement and educational impact. 

This short paper presents a {\em reference architecture} for gamified cultural heritage applications leveraging {\em generative AI} and {\em augmented reality}. {\em Gamification} enhances motivation, {\em artificial intelligence} enables adaptive storytelling and personalized content, and {\em augmented reality} fosters immersive, location-aware experiences. Integrating AI with gamification supports dynamic mechanics, personalized feedback, and user behavior prediction, improving engagement. The modular design supports {\em scalability}, {\em interoperability}, and {\em adaptability} across heritage contexts. 

This research provides a framework for designing {\em interactive} and {\em intelligent} cultural heritage applications, promoting accessibility and deeper appreciation among users and stakeholders.

\end{abstract}
%
%
\section{Introduction and motivation}

The rapid advancements in Information and Communication Technologies (ICT) are transforming access to and interaction with cultural heritage (CH) \cite{sylaiou2019ict}. The rise of mixed reality devices, such as smart glasses, marks a shift from conventional mobile applications to immersive, context-aware experiences, aligning with the growing demand for interactive and personalized CH engagement \cite{bekele2018survey}.

Over the past decade, mixed reality has gained traction in serious games and gamification (GAM), particularly in education \cite{zikas2016mixed} and CH \cite{ioannides2017mixed}. Meanwhile, Generative Artificial Intelligence (GAI) is revolutionizing digital content creation by enabling adaptive storytelling, automated translation, and real-time personalization \cite{casillo2024role}.

Despite these advancements, no unified framework currently integrates GAM, GAI and Augmented Reality (AR) to enhance CH experiences. While gamification boosts engagement \cite{khan2022role}, GAI enables dynamic content generation and intelligent adaptation \cite{erbacsi2023role} and AR provides interactive, spatially aware visualizations \cite{bekele2018survey}, their combined potential remains largely unexploited.

This paper introduces a \textit{reference software architecture} for next-generation CH applications, merging gamification, AI-driven generative content and AR within a scalable, adaptive framework. Enabling real-time content generation, fostering engagement, and supporting immersive interactions, this architecture lays the foundation for intelligent, responsive CH experiences and provides a structured blueprint for future research into AI-driven personalization, multimodal interaction, and the integration of emerging technologies.

The paper is structured as follows: Section \ref{Section2} reviews related work, identifying research gaps and existing architectures. Section \ref{Section3} details the proposed architecture, emphasizing its modularity, scalability, and integration strategies. Finally, Section \ref{Section4} summarizes key contributions and outlines future research directions.

\section{Related Work} \label{Section2}

The convergence of GAM, GAI, and AR is reshaping CH dissemination and education. Despite individual advancements in each field, their seamless integration within a unified architectural framework remains an open challenge. This section reviews existing approaches, identifies research gaps, and highlights the need for a structured reference architecture that effectively combines these technologies in CH applications.

{\em Gamification} enhances user engagement and motivation in CH experiences \cite{ccetin2021gamification}. Marques et al. \cite{marques2023systematic} emphasize its role in fostering creativity but note the absence of standardized architectural frameworks. Similarly, Khan et al. \cite{khan2022role} highlight the lack of structured design methodologies, leading to fragmented, non-scalable implementations. While game mechanics such as challenges, rewards, and progress tracking through levels are commonly employed, their integration into a modular, adaptable architecture for CH remains underexplored.

{\em Augmented Reality} has revolutionized CH storytelling and visualization through immersive digital overlays on historical artifacts and sites. However, many research prototypes lack a strong architectural foundation \cite{reicher2003results}. Panou et al. \cite{panou2018architecture} propose a mobile AR-based tourist guide incorporating gamification, addressing challenges like real-time tracking and content synchronization. Yet, most AR systems rely on marker- or location-based tracking without adaptive content delivery, limiting their ability to dynamically personalize user experiences. The integration of {\em AI} for real-time adaptation, contextual storytelling, and interactive dialogue generation remains underdeveloped.

The combined potential of {\em Gamification, AI} and {\em AR} for intelligent CH experiences is evident, but a comprehensive reference architecture unifying these elements is still lacking. Ribeiro et al. \cite{ribeiro2024vr} explore an AI-driven, VR-AR gamified museum system that improves engagement and accessibility. However, their work does not propose a reference software architecture applicable across diverse CH contexts. AI-driven models can personalize narratives, optimize user interactions, and generate dynamic storytelling using machine learning (ML) and natural language processing (NLP). Current solutions mainly focus on standalone recommendation systems rather than fully integrated {\em GAM-AI-AR} frameworks.

Despite progress in individual and combined applications, the absence of a {\em cohesive, scalable, and modular} {\em reference architecture} remains a key barrier to developing integrated, adaptive, and interoperable CH systems. To address this, the present work introduces a {\em novel reference architecture} that systematically integrates these technologies as described in the next section.

\section{The Conceptual Reference Architecture} \label{Section3}

A {\em reference architecture} defines the core subsystems and their interactions within a domain, providing a standardized framework for system design and evolution. It enhances development and maintenance by improving system comprehension, supporting trade-off analysis, and serving as a structured template for designing or refining systems \cite{cloutier2010concept}. Establishing a common architectural foundation should ensure {\em scalability, interoperability, and adaptability} to evolving technological and user needs \cite{reicher2003results,ribeiro2024vr,siriwardhana2021survey}.

This section introduces the {\em Conceptual Reference Architecture} (Fig. \ref{fig:refarch}). The architecture integrates {\em gamification mechanics, AI-powered recommendations, and AR visualization} to enhance engagement, optimize performance, and enable seamless real-time interactions within a scalable ecosystem.

To meet these objectives, the architecture aligns with key {\em Non-Functional Requirements (NFRs)}, such as scalability, interoperability, real-time performance, and security, ensuring smooth integration across CH institutions and digital platforms. It also incorporates {\em Key Design Options (KDOs)}, including event-driven communication, hybrid cloud-edge processing, and modular service orchestration, providing flexibility and efficiency \cite{hossain2023role}.

The following subsections detail the identified NFRs, KDOs and the core components of the architecture, their interactions, and the mechanisms enabling real-time adaptation and personalized user experiences for CH applications.



\subsection{Non-Functional Requirements}

Developing a scalable and cohesive software architecture that integrates GAM, GAI, and AR for CH applications requires careful architectural considerations. The integration of these technologies must ensure {\em maintainability}, {\em adaptability}, {\em interoperability}, {\em scalability}, {\em performance}, {\em modularity}, and {\em personalization}, balancing real-time \textit{responsiveness} with the complexity of AI-driven content generation and adaptive user experiences. This section outlines the identified essential {\em Non-Functional Requirements (NFR)} necessary to achieve an efficient and sustainable architecture.

To enhance {\em maintainability and adaptability}, the architecture must adhere to modular design principles, ensuring a clear separation between AI-driven recommendations, gamification mechanics, and AR visualization. This modular approach allows independent component updates without disrupting system functionality. Moreover, {\em interoperability and integration} play a crucial role in enabling CH institutions to share digital assets and integrate external datasets. Supporting open data standards (e.g., GLTF, X3D, RDF, OWL) and providing flexible API-based communication fosters seamless content exchange.

{\em Scalability} is a fundamental requirement, as CH applications must accommodate an increasing number of users and collections, dynamic AI-generated content, and complex AR renderings \cite{reicher2003results} and interactions while maintaining system {\em performance and responsiveness}. The architecture should support cloud-edge hybrid processing to offload AI inference and optimized AR rendering, while load balancing and event-driven mechanisms ensure resource efficiency \cite{siriwardhana2021survey}.

{\em Personalization} is essential for user engagement, requiring AI-driven adaptive storytelling, dynamic challenge adjustments, and personalized AR content placement based on user behavior. AI should enable context-aware recommendations and real-time content adaptation, ensuring tailored cultural exploration experiences. Furthermore, {\em usability and accessibility} must be prioritized by implementing multimodal interactions (voice, touch, gesture), adaptive user interface scaling, and compliance with accessibility standards (WCAG 2.1) \cite{spina2019wcag}.

Given the reliance on personalized AI-driven content, {\em security and privacy} mechanisms are critical. Secure authentication, anonymized data tracking, and AI explainability mechanisms must be embedded to ensure ethical and transparent AI usage. Additionally, {\em reliability and fault tolerance} must be addressed through offline support, resilient AR tracking, and self-healing infrastructure mechanisms, ensuring uninterrupted functionality in diverse museum and heritage site environments. To maintain {\em data integrity}, historical narratives and AI-generated content should be validated by domain experts, stored in version-controlled knowledge bases, and monitored for factual accuracy.

\subsection{Key Design Options}

To implement these NFRs, specific KDOs must be adopted. A {\em hybrid microservices and event-driven architecture} ensures AI, AR, and gamification components operate independently while synchronizing in real time \cite{rahmatulloh2022event}. As users interact with AR elements and perform game actions, an event-driven mechanism updates game status and dynamically adjust rewards. Message queues and WebSockets facilitate smooth synchronization between AI recommendations, AR overlays, and game progress.

A {\em modular architecture} decouples GAM services, AI and AR. The Gamification Engine manages engagement tracking and adaptive difficulty, while the GAI module employs ML and LLMs to generate personalized narratives and exploration pathways \cite{trichopoulos2023large}. The AR Module ensures real-time rendering, spatial tracking, and multimodal interaction for an immersive {\em phygital} experience. 

For seamless integration, RESTful APIs support structured data exchange, WebSockets enable low-latency interactions, and message queues handle asynchronous AI content processing. {\em Hybrid data management} employs Structured Query Language (SQL) for structured user tracking and NoSQL for AI-generated content. Content Delivery Networks (CDNs) optimize AR content distribution, reducing rendering latency.

As AI-driven personalization is core to CH applications, reinforcement learning, context-aware AR placement, and adaptive gamification should be integrated. AI analyzes user behavior and real-time interactions to refine recommendations while ensuring GDPR-compliant data handling.

Finally, {\em future extensibility} is enabled by containerization, plugin-based AI, and decoupled training and inference pipelines. This ensures continuous AI updates, integration of new gamification mechanics, and adaptation to emerging interaction paradigms, supporting long-term evolution in CH applications.


\subsection{Main Components and their Integration}

\begin{figure}[h!]
    \centering
    \includegraphics[width=0.9\textwidth]{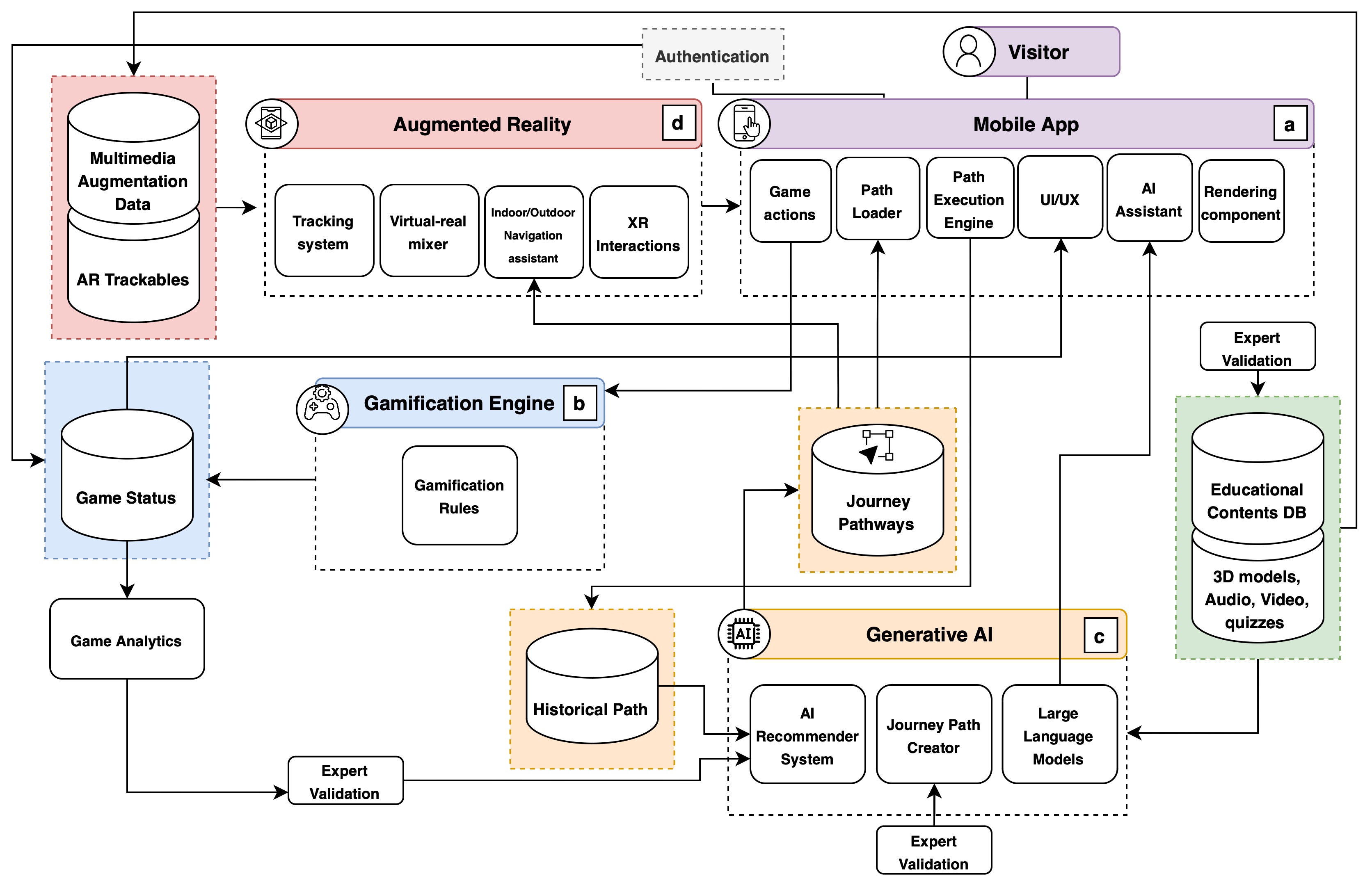}
    \caption{The proposed Reference Architecture.}
    \label{fig:refarch}
\end{figure}

The proposed {\em reference architecture} (Figure \ref{fig:refarch}) integrates {\em gamification, AI-driven personalization}, and {\em augmented reality} to enhance cultural heritage experiences. This architecture defines core components that work together to create interactive, immersive, and adaptive user experiences\footnote{For the sake of space, this short paper will cover the component view only, while leaving further views to future work.}.

\begin{figure}[htb]
    \centering
    \includegraphics[width=1\textwidth]{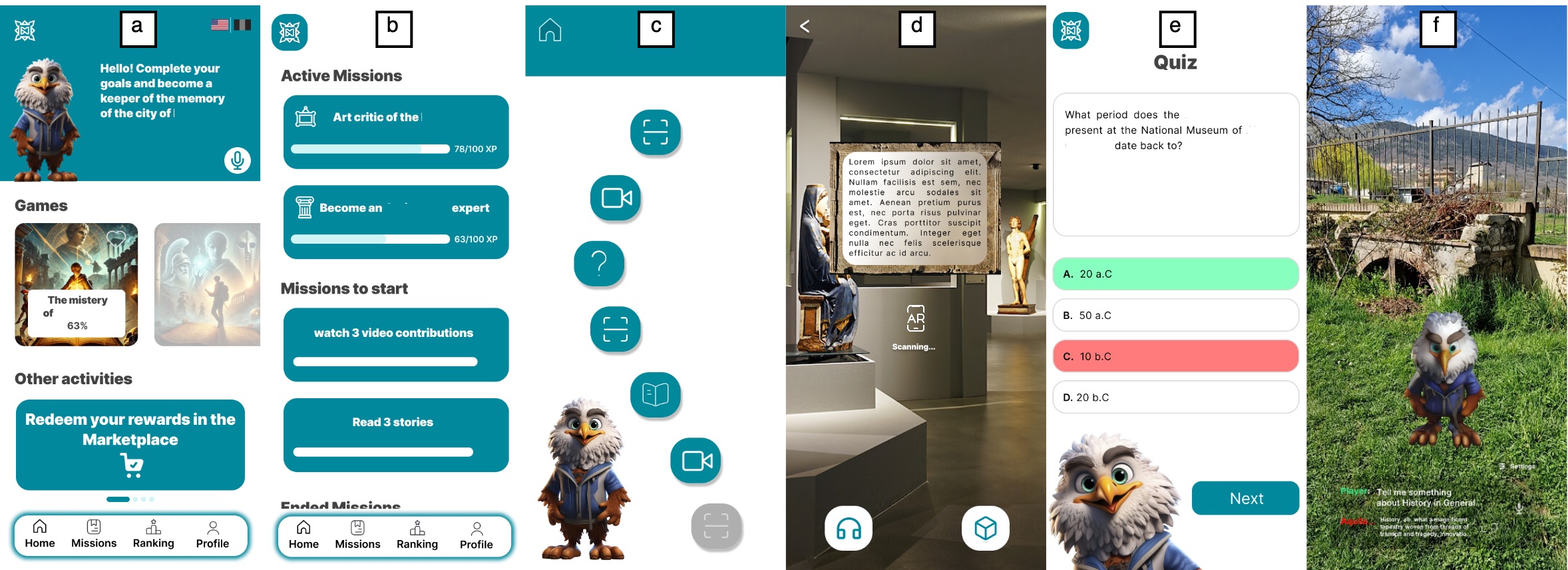}
    \caption{Screenshots from the App. From left to right: The Home Page, the game status page, the journey page, the AR visualizations of artifacts, an example of a Quiz and the AR AI assistant.}
    \label{fig:MobileApp}
\end{figure}

At the heart of the system is the \textit{Mobile App} (Figure \ref{fig:refarch}.a), which serves as the primary interface for user interaction, content delivery, and engagement. It structures CH visits into adaptive exploration journeys, guiding users through dynamically generated pathways that integrate AR content and gamified challenges. The \texttt{Path Execution Engine} interprets user-specific journeys as graph-based structures, where each node of the graph represent a Point of Interest (POI) and sequentially unlock as users engage with content and challenges. Progression through this graph is determined by user interactions, AI-driven recommendations, and gamification scores. Each POI integrates AR visualizations, interactive storytelling, and game-based rewards, ensuring a structured yet adaptive cultural heritage exploration experience (Figure \ref{fig:MobileApp}.c), while the \texttt{AI Assistant} provides historical insights, personalized recommendations, and interactive storytelling (Figure \ref{fig:MobileApp}.f). The \texttt{Rendering Component} processes and displays AR elements, overlaying digital content onto real-world artifacts to enhance immersion while enabling interaction with 3D objects (Figure \ref{fig:MobileApp}.d). It supports real-time manipulation, scaling, and rotation of virtual reconstructions, allowing users to explore historical artifacts from multiple perspectives.

\sloppy The {\em Gamification Engine} (Figure \ref{fig:refarch}.b) plays a central role in driving engagement by managing game logic, tracking progress, and executing challenges. It integrates elements such as points, badges, levels, and leaderboards, ensuring that users remain motivated throughout their exploration. The engine processes \texttt{game actions} triggered by user interactions, evaluating them against predefined rules and updating the \texttt{game status} in real time (Figure \ref{fig:MobileApp}.b). A {\em Domain-Specific Language (DSL)} \cite{BucchiaroneMFM23} facilitates the creation of \texttt{gamification rules}, enabling structured and adaptable game mechanics. 
The engine also aggregates engagement data, including user interactions with AR objects, quizzes, and challenges, and forwards this data to the \texttt{AI Recommender System}. The AI Recommender analyzes this information—alongside user preferences and exploration history—to determine the optimal next POI within the graph-based journey structure.

A key enabler of personalization is the \textit{Generative AI module} (Figure \ref{fig:refarch}.c), which dynamically generates content, optimizes exploration pathways, and enables AI-assisted storytelling (Figure \ref{fig:MobileApp}.f). It leverages ML, NLP, and recommendation algorithms to tailor each visitor’s journey. The \texttt{AI Recommender System} analyzes user behavior to suggest personalized exploration paths modeled as directed graph, where each node unlocks specific activities, challenges, and rewards. The \texttt{Journey Path Creator} dynamically adjusts the traversal of this graph based on user preferences, past interactions, and real-time engagement, ensuring an adaptive and interactive exploration. Each node is linked to predefined gamification mechanics, influencing progress through AI-driven recommendations and contextual AR storytelling. To ensure historical accuracy, AI-generated content undergoes \texttt{expert validation} before being added to the \texttt{Educational Content Database}, allowing continuous refinement through expert feedback and user interactions.

\sloppy
Enhancing the phygital interaction, the \textit{Augmented Reality System} (Figure \ref{fig:refarch}.d) blends virtual and real-world elements, creating an immersive cultural experience. The \texttt{Multimedia Augmentation Data Repository} stores AR trackables, 3D models, and digital overlays, while the \texttt{tracking system} ensures precise positioning using spatial recognition, GPS localization, or marker-based tracking. Users can access AR content through various devices, including smartphones, tablets, and smart glasses. The \texttt{virtual-real mixer} integrates digital elements into historical sites, presenting interactive reconstructions, animated narratives, and contextual information. Additionally, the \texttt{indoor/outdoor navigation assistant} provides real-time guidance based on the user’s journey, while \texttt{XR interactions} enable multimodal engagement through touch, voice, and gestures (Figure \ref{fig:MobileApp}.d,f).


\section{Conclusion and future directions} \label{Section4}

The proposed architecture originates from a prototype under development for \textit{The National Museum of \footnote{The name was removed for double-blind review.}}, designed to enhance visitor engagement and promote lesser-known archaeological sites. By leveraging event-driven mechanisms and hybrid cloud-edge processing, the architecture proposes real-time responsiveness and seamless integration of {\em gamification}, {\em AI personalization}, and {\em AR-based exploration}.

Its structured design enables interoperability with external CH repositories and datasets, ensuring long-term adaptability. AI-assisted journey pathways and dynamic AR overlays enhance context-aware exploration, fostering engagement and knowledge retention. Human-in-the-Loop validation guarantees historical accuracy in AI-generated content.

Future work will refine AI-driven personalization, optimize component synchronization, and conduct extensive \textit{validation and testing}. User trials will assess usability, engagement, and educational impact, guiding iterative improvements for broader adoption.

%
%
%
 \bibliographystyle{splncs04}

\begin{thebibliography}{10}
\providecommand{\url}[1]{\texttt{#1}}
\providecommand{\urlprefix}{URL }
\providecommand{\doi}[1]{https://doi.org/#1}

\bibitem{BucchiaroneMFM23}
Removed for double-blind review

\bibitem{bekele2018survey}
Bekele, M.K., Pierdicca, R., Frontoni, E., Malinverni, E.S., Gain, J.: A survey of augmented, virtual, and mixed reality for cultural heritage. Journal on Computing and Cultural Heritage (JOCCH)  \textbf{11}(2),  1--36 (2018)

\bibitem{casillo2024role}
Casillo, M., Colace, F., Gupta, B.B., Lorusso, A., Santaniello, D., Valentino, C.: The role of ai in improving interaction with cultural heritage: An overview. Handbook of Research on AI and ML for Intelligent Machines and Systems  (2024)

\bibitem{ccetin2021gamification}
{\c{C}}etin, {\"O}., Erbay, F.: Gamification practices in museums. Journal of Tourismology  \textbf{7}(2),  265--276 (2021)

\bibitem{cloutier2010concept}
Cloutier, R., Muller, G., Verma, D., Nilchiani, R., Hole, E., Bone, M.: The concept of reference architectures. Systems Engineering  \textbf{13}(1),  14--27 (2010)

\bibitem{erbacsi2023role}
Erba{\c{s}}{\i}, Z., Tural, B., {\c{C}}o{\c{s}}kuner, {\.I}.: The role and potential of artificial intelligence and gamification in education: The example of vak{\i}f participation bank. Orclever Proc. of Research and Development  \textbf{3}(1),  243--254 (2023)

\bibitem{hossain2023role}
Hossain, M.D., Sultana, T., Akhter, S., Hossain, M.I., Thu, N.T., Huynh, L.N., Lee, G.W., Huh, E.N.: The role of microservice approach in edge computing: Opportunities, challenges, and research directions. ICT Express  \textbf{9}(6),  1162--1182 (2023)

\bibitem{ioannides2017mixed}
Ioannides, M., Magnenat-Thalmann, N., Papagiannakis, G.: Mixed reality and gamification for cultural heritage, vol.~2. Springer (2017)

\bibitem{khan2022role}
Khan, I., Melro, A., Amaro, A.C., Oliveira, L.: Role of gamification in cultural heritage dissemination: A systematic review. In: Proc. of Sixth Int. Congress on Information and Communication Technology: ICICT 2021, London, Volume 1. pp. 393--400. Springer (2022)

\bibitem{marques2023systematic}
Marques, C.G., Pedro, J.P., Ara{\'u}jo, I.: A systematic literature review of gamification in/for cultural heritage: leveling up, going beyond. Heritage  \textbf{6}(8) (2023)

\bibitem{panou2018architecture}
Panou, C., Ragia, L., Dimelli, D., Mania, K.: An architecture for mobile outdoors augmented reality for cultural heritage. ISPRS Int. Journal of Geo-Information  \textbf{7}(12), ~463 (2018)

\bibitem{rahmatulloh2022event}
Rahmatulloh, A., Nugraha, F., Gunawan, R., Darmawan, I.: Event-driven architecture to improve performance and scalability in microservices-based systems. In: 2022 Int. Conference Advancement in Data Science, E-learning and Information Systems (ICADEIS). pp. 01--06. IEEE (2022)

\bibitem{reicher2003results}
Reicher, T., Mac~Williams, A., Brugge, B., Klinker, G.: Results of a study on software architectures for augmented reality systems. In: The Second IEEE and ACM Int. Symposium on Mixed and Augmented Reality. Proc. IEEE (2003)

\bibitem{ribeiro2024vr}
Ribeiro, M., Santos, J., Lobo, J., Ara{\'u}jo, S., Magalh{\~a}es, L., Ad{\~a}o, T.: Vr, ar, gamification and ai towards the next generation of systems supporting cultural heritage: Addressing challenges of a museum context. In: Proc. of the 29th Int. ACM Conference on 3D Web Technology. pp. 1--10 (2024)

\bibitem{siriwardhana2021survey}
Siriwardhana, Y., Porambage, P., Liyanage, M., Ylianttila, M.: A survey on mobile augmented reality with 5g mobile edge computing: Architectures, applications, and technical aspects. IEEE Communications Surveys \& Tutorials  \textbf{23}(2) (2021)

\bibitem{spina2019wcag}
Spina, C.: Wcag 2.1 and the current state of web accessibility in libraries. Weave: Journal of Library User Experience  \textbf{2}(2) (2019)

\bibitem{sylaiou2019ict}
Sylaiou, S., Papaioannou, G.: Ict in the promotion of arts and cultural heritage education in museums. In: Strategic Innovative Marketing and Tourism: 7th ICSIMAT, Athenian Riviera, Greece, 2018. pp. 363--370. Springer (2019)

\bibitem{trichopoulos2023large}
Trichopoulos, G.: Large language models for cultural heritage. In: Proc. of the 2nd Int. Conference of the ACM Greek SIGCHI Chapter. pp.~1--5 (2023)

\bibitem{zikas2016mixed}
Zikas, P., Bachlitzanakis, V., Papaefthymiou, M., Kateros, S., Georgiou, S., Lydatakis, N., Papagiannakis, G.: Mixed reality serious games and gamification for smart education. In: European conference on games based learning. p.~805. Academic Conferences Int. Limited (2016)

\end{thebibliography}

%






\end{document}